\documentclass[aps,prl,floatfix,twocolumn]{revtex4}
\usepackage{graphicx,amsmath,units,xspace}

\graphicspath{{figures/}}

\newcommand{\micron}{\ensuremath{\unit{\mu m}}\xspace}
\newcommand{\order}[1]{\ensuremath{{\mathcal O}(#1)}\xspace}
\newcommand{\Th}[1]{\ensuremath{T_\text{h}^{(#1)}}\xspace}
\newcommand{\Ths}{\Th{s}}
\newcommand{\Tcon}[1]{\ensuremath{T_\text{con#1}}\xspace}
\newcommand{\TconF}{\ensuremath{T_\text{conF}}\xspace}
\renewcommand{\vec}[1]{\ensuremath{{\mathbf #1}}\xspace}

\begin{document}

\title{Configurational Temperature of Charge-Stabilized Colloidal Monolayers}

\author{Yilong Han}

\author{David G. Grier}

\affiliation{Dept.~of Physics, James Franck Institute 
  and Institute for Biophysical Dynamics \\
  The University of Chicago, Chicago, IL 60637}

\date{\today}

\begin{abstract}
Recent theoretical advances show that the temperature of a system in equilibrium
can be measured
from static snapshots of its constituents' instantaneous configurations, without
regard to their dynamics.
We report the first measurements of the configurational temperature in
an experimental system.
In particular, we introduce a hierarchy of hyperconfigurational temperature
definitions, which we use to analyze
monolayers of
charge-stabilized colloidal spheres.
Equality of the hyperconfigurational and bulk thermodynamic temperatures provides
previously lacking thermodynamic self-consistency checks for the measured
colloidal pair potentials, and thereby casts new light on  
anomalous like-charge colloidal attractions induced by
geometric confinement.
\end{abstract}

% 82.70.Dd Colloids
% 05.40.-a Thermodynamics
% 61.20.p Structure of Liquids

\maketitle

The temperature of an equilibrium ensemble of particles
is defined conventionally in terms of the particles' mean kinetic energy, 
without regard for their instantaneous positions.
In 1997, Rugh pointed out that the temperature
can also emerge from other ensemble averages over geometrical
and dynamical quantities \cite{rugh97}.
This notion is expressed more generally \cite{jepps00,rickayzen01} as
\begin{equation}
  k_B T = \frac{\langle \nabla {\cal H} \cdot \vec{B}(\vec{\Gamma})\rangle}{%
    \langle \nabla \cdot \vec{B}(\vec{\Gamma}) \rangle},
  \label{eq:general}
\end{equation}
where angle brackets indicate an ensemble (or time) average,
$\vec{\Gamma} = \{q_1, \dots, q_{3N}, p_1, \dots, p_{3N}\}$ is the
instantaneous set of $3N$ generalized coordinates $q_j$ and their
conjugate momenta $p_j$ for an $N$-particle system, 
${\cal H} = \sum_{j = 1}^{3N} p_j^2 / (2m) + V(\{q_j\})$ is the
Hamiltonian associated with the conservative $N$-particle potential
$V(\{q_j\})$, and $\vec{B}(\vec{\Gamma})$ is an \emph{arbitrary} vector field
selected so that both the numerator and denominator of Eq.~(\ref{eq:general})
are finite and the numerator grows more slowly than $e^N$ in
the thermodynamic limit.
Choosing
$\vec{B}(\vec{\Gamma}) = \{0, \dots,0, p_1, \dots p_{3N}\}$ yields the
usual kinetic definition of temperature.
Choosing instead $\vec{B}(\vec{\Gamma}) = - \nabla V(\{q_i\})$
yields a formally equivalent result,
\begin{equation}
  k_B T_\text{config} = \frac{\langle \left|\nabla V\right|^2 \rangle}{%
    \langle \nabla^2 V \rangle},
  \label{eq:tconfig}
\end{equation}
which depends only on the particles' instantaneous configuration,
and not on their momenta.
Variants of Eq.~(\ref{eq:tconfig}) have been widely
adopted in molecular dynamics simulations \cite{butler98,delhommelle02},
but have not previously been applied to experiments.

In this Letter, we use the configurational temperature formalism to
probe macroionic interactions in
monolayers of charged colloidal spheres dispersed in water and 
confined between parallel glass plates.
These measurements provide sensitive self-consistency tests for the
measured inter-particle pair potentials,
and thereby provide new insights into the long-standing conundrum of anomalous
attractions in geometrically confined charge-stabilized dispersions.

Directly applying Eq.~(\ref{eq:tconfig}) requires the
full $N$-particle free energy, which is rarely available.
Simplified forms emerge for systems satisfying certain conditions.
For example, if $V(\{q_i\})$ is
the linear superposition of pair potentials, $u(r)$, then
Eq.~(\ref{eq:tconfig}) reduces to \cite{butler98},
\begin{equation}
  k_B \TconF = - \frac{\langle \sum_{i=1}^N F_i^2 \rangle}{%
    \langle \sum_{i=1}^N \nabla_i \cdot \vec{F}_i\rangle},
\label{eq:TconF}
\end{equation}
where $\vec{F}_i = -\sum_{j\ne i} \nabla_i u(r_{ij})$ 
is the total force on particle $i$ due to its
interactions with other particles, $\nabla_i$ is the gradient with
respect to the $i$-th particle's position, $\vec{r}_i$,
and $r_{ij} = |\vec{r}_i - \vec{r}_j|$ is the center-to-center
separation between particles $i$ and $j$.
The temperature is reflected in the instantaneous distribution
of forces because objects explore more of 
their potential energy landscape as the temperature increases.

Equation~(\ref{eq:TconF}) may be generalized into a hierarchy of
\emph{hyperconfigurational} temperatures
by choosing $\vec{B}(\vec{\Gamma}) = \{F_i^s\}$:
\begin{equation}
  k_B \Ths = - \frac{\langle \sum_{i=1}^N F_i^{s+1} \rangle}{%
    \langle s \sum_{i=1}^N F_i^{s-1} \, \nabla_i \cdot \vec{F}_i \rangle},
  \label{eq:thyper}
\end{equation}
for $s > 0$.
These higher moments are more sensitive to
the input potential's detailed structure than $\TconF = \Th{1}$.
They also can be applied to three-dimensional systems with long-ranged $1/r$ potentials,
for which \TconF is ill-defined.

The same hierarchy also may be derived from the classical hypervirial theorem
\cite{hirschfelder60},
$\langle \{f,{\mathcal H}\} \rangle = 0$, where $\{\cdots\}$ is the Poisson bracket,
by selecting $f(\vec{\Gamma}) = \sum_{j=1}^N p_j \, F_j^s.$ 
More generally, $f(\vec{\Gamma})$ can be any finite-valued function that does not
explicitly depend on time.

For systems with short-ranged potentials, dropping additional terms of $\order{1/N}$ from 
Eq.~(\ref{eq:general}) yields \cite{jepps00}:
\begin{align}
  k_B \Tcon{1} & = - \left< \frac{\sum_{i = 1}^N F_i^2}{%
    \sum_{i = 1}^N \nabla_i \cdot \vec{F}_i} \right>, \quad \text{and} \label{eq:tcon1}\\
  k_B \Tcon{2} & =  - \left< \frac{\sum_{i = 1}^N \nabla_i \cdot \vec{F}_i}{%
    \sum_{i = 1}^N F_i^2} \right>^{-1}, \label{eq:tcon2}
\end{align}
the second of which is proposed here for the first time.
These definitions' different dependences on sample size $N$ are useful for
comparison with \Ths.

Temperature definitions based on configurational information are
ideal for studying colloidal spheres dispersed in
viscous fluids such as water.
The fluid acts as a heat bath at temperature $T$.
It also randomizes the particles' motions over intervals longer than the viscous relaxation
time, typically measured in microseconds.
As a result, the colloids' instantaneous momenta are not easily accessible.
Their positions, however, are readily measured using standard methods
of digital video microscopy \cite{crocker96}.
Calculating the temperature from the resulting positional data then requires
accurate knowledge of the colloids' interactions.

The mean-field theory for macroionic interactions \cite{russel89}
predicts that charge-stabilized colloidal spheres should repel
each other through a screened-Coulomb potential:
\begin{equation}
  \label{eq:dlvo}
  \beta u(r) = Z^2 \lambda_B \, \frac{\exp(\kappa \sigma)}{
    \left(1 + \frac{\kappa \sigma}{2}\right)^2} \, \frac{\exp(-\kappa r)}{r},
\end{equation}
where $\beta^{-1} = k_B T$ is the heat bath's thermal energy scale,
$\sigma$ is the spheres' diameter, $r$ is their center-to-center separation, 
$Z$ is the effective charge number on each sphere and $\lambda_B = e^2/(4\pi \epsilon k_B T)$ 
is the Bjerrum length in a medium of dielectric constant $\epsilon$.
The Debye-H\"uckel screening length $\kappa^{-1}$ sets the range of 
the effective electrostatic interaction and depends on the concentration $c$ of
(monovalent) ions through $\kappa^2 = 4 \pi c \lambda_B$.

\begin{figure}[t!]
  \centering
  \includegraphics[width=\columnwidth]{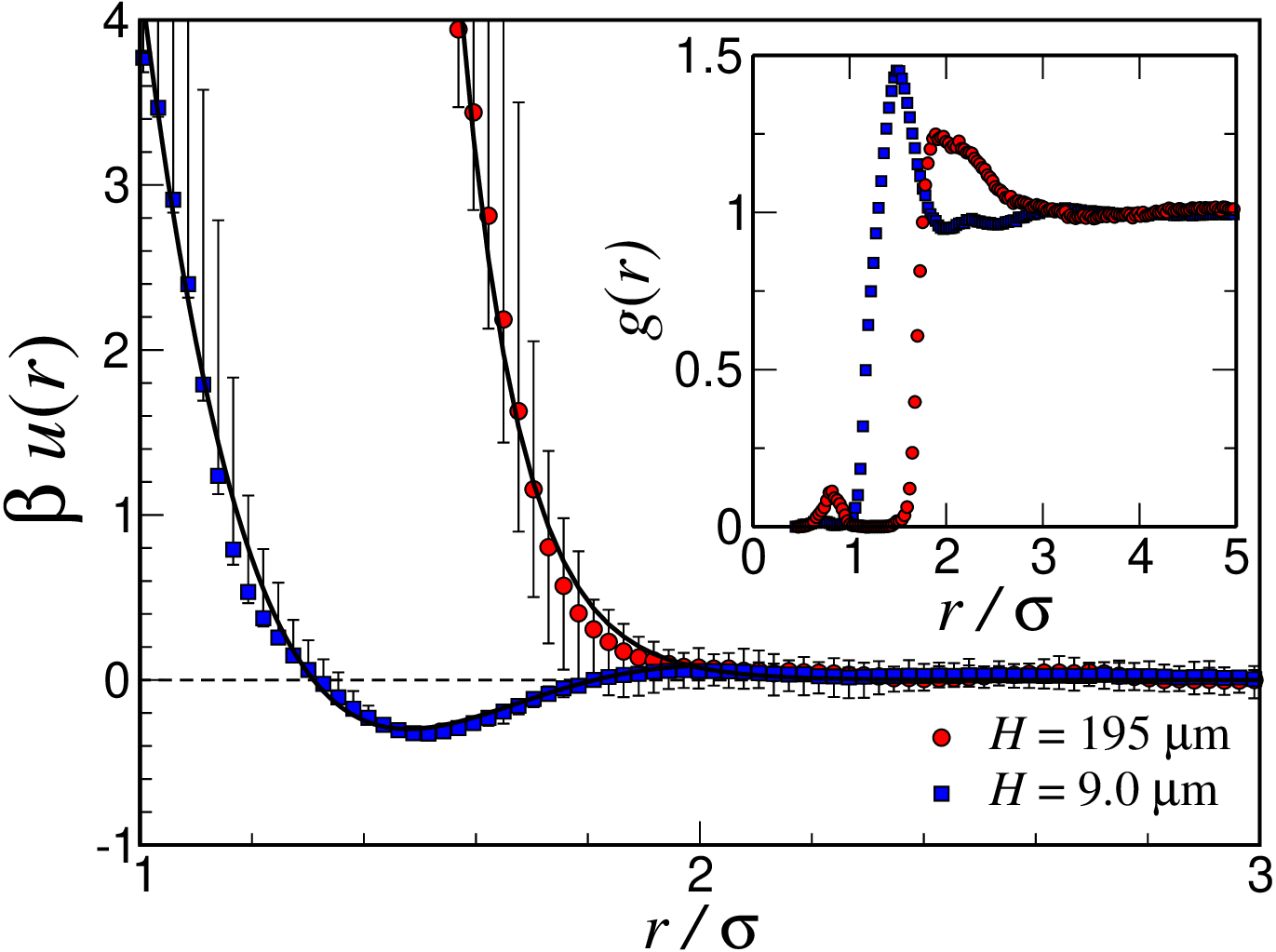}
  \caption{Pair potentials $u(r)$ for silica colloid $\sigma = 1.58~\micron$
    in diameter extracted from the measured radial distribution
    functions $g(r)$ using the HNC approximation for suspensions at
    $H = 9~\micron$, $n\sigma^2 = 0.0654$ (squares) and
    $H = 195~\micron$, $n\sigma^2 = 0.0797$ (circles). 
    Solid curves are, respectively, fits to a 5-th order polynomial and 
    to Eq.~(\protect\ref{eq:dlvo})
    ($\kappa^{-1} = 180 \pm 10~\unit{nm}$, $Z = 6500 \pm 1000$).
}
  \label{fig:ur}
\end{figure}

Despite its success at explaining bulk colloidal phenomena \cite{russel89},
mean-field theory qualitatively fails \cite{meanfield}
to explain the strong and long-ranged 
attractions observed when charged spheres are confined between parallel glass walls
\cite{kepler94,crocker96,carbajaltinoco96,han03a}.
The crossover from monotonic repulsion to long-ranged attraction with increasing
confinement is demonstrated in Fig.~\ref{fig:ur}.
Efforts to explain this anomaly through non-mean-field mechanisms 
so far have not yielded the experimentally observed  effect \cite{gopinathan02}. 
Given the apparent difficulty of explaining anomalous attractions on the basis 
of colloidal electrostatics or electrodynamics, various other explanations 
have been proposed, most of which focus on possible experimental artifacts.
For example, nonequilibrium hydrodynamic coupling has been shown \cite{squires00} 
to explain one measurement based on optical tweezer manipulation \cite{larsen97},
but cannot be relevant for measurements
based on video microscopy of colloid in equilibrium 
\cite{kepler94,carbajaltinoco96,vondermassen94,behrens01a,han03a}.
Concern also has been raised that such imaging measurements
can fall victim to correlated artifacts due to
statistical fluctuations \cite{behrens01a}.  
We have shown, however, that anomalous attractions still are clearly
resolved given adequate statistics \cite{han03a}.
A still greater
concern is that uncorrected many-body artifacts could mimic attractions
in purely repulsive dispersions.
Such artifacts certainly can arise at high densities, as has
been demonstrated experimentally \cite{brunner02}.  
Evidence for pairwise additivity at lower densities
has relied principally on comparisons over a range
of concentrations, for which variations in chemical environment could mask other effects.

We resolve all such ambiguities by exploiting the configurational temperatures' sensitivity to
inaccuracies in the input potential as self-consistency tests for colloidal interaction 
measurements.
In particular, we use pair potentials measured according to
Refs.~\cite{crocker96}, \cite{han03a} and \cite{behrens01a} to compute the configurational
temperature of geometrically confined colloidal monolayers and compare the results
with the bulk thermodynamic temperature, with $\Ths = T$ signaling thermodynamic
self-consistency.
Deviations could result from a breakdown of pairwise additivity in a system
with non-trivial many-body interactions, a departure from equilibrium in a system subjected to 
hydrodynamic forces, or simply inaccuracy in $u(r)$.

Our samples consist of uniform silica spheres
$\sigma = 1.58 \pm 0.03~\micron$ in diameter (Duke Scientific Lot 24169) dispersed in deionized water
and loaded into hermetically sealed sample volumes
formed by bonding the edges of glass \#1.5 coverslips to the surfaces of glass
microscope slides.
The separation $H$ between these surfaces is set during construction and establishes the
degree of confinement.
Reservoirs of mixed bed ion exchange resin maintain the ionic strength below
$c \approx 10~\unit{\mu M}$, corresponding to a Debye-H\"uckel screening length
$\kappa^{-1} \gtrsim 150~\unit{nm}$.
Under these conditions, both the silica spheres and the glass walls develop negative 
surface charge densities of roughly $10^3 \, e_0~\unit{\micron^{-2}}$, where $e_0$ is
the electron charge \cite{behrens01b}.
Samples are mounted on the stage of a Zeiss Axiovert 100 STV microscope,
after which
the dense silica spheres rapidly sediment into a monolayer roughly
$h = 900~\unit{nm}$ above the coverslip with out-of-plane fluctuations smaller
than 300~\unit{nm}.
The bright-field imaging system provides a magnification of 212~\unit{nm/pixel}
on a Hitachi TI-11A monochrome CCD camera.
The resulting video stream is recorded and digitized into 60 deinterlaced video
fields per second, each of which is analyzed to yield the instantaneous distribution
\begin{equation}
  \label{eq:distribution}
  \rho(\vec{r}) = \sum_{i = 1}^{N(t)} \delta(\vec{r} - \vec{r}_i(t))
\end{equation}
of $N(t)$ spheres at locations $\vec{r}_i(t)$ at time $t$.
Measurements were performed at room temperature over periods of roughly one hour, with temperature
fluctuations smaller than $\pm 1~\unit{K}$ over the course of any measurement.

From $\rho(\vec{r},t)$, we calculate the
radial distribution function,
\begin{equation}
  \label{eq:gr}
  g(r) = \frac{1}{n^2} \, \langle \rho(\vec{r}^\prime - \vec{r}, t) \, 
  \rho(\vec{r}^\prime,t) \rangle,
\end{equation}
where $n = N/A$ is the areal density of $N = \langle N(t) \rangle$ particles in 
the field of view of area $A$, and the angle brackets indicate averages over angles and time.
Assuming that the system is in equilibrium and that its interactions are isotropic and pairwise
additive,
$g(r)$ may be inverted to obtain an estimate for $u(r)$ \cite{kepler94,carbajaltinoco96}.
At low enough densities, the Boltzmann distribution provides the necessary relationship:
$\beta u(r) = - \lim_{n \rightarrow 0} \ln g(r)$.
Unfortunately, the need to sample $g(r)$ adequately in a limited field of view requires
higher concentrations \cite{behrens01a,han03a}, with $0.05 < n \sigma^2 < 0.1$
being typical for our experiments.
We extract candidate pair potentials $u(r)$ such as the examples in Fig.~\ref{fig:ur} using
the liquid-structure inversion method based on the Ornstein-Zernike equation 
with hypernetted-chain (HNC) and Percus-Yevick (PY) closures \cite{carbajaltinoco96,behrens01a,han03a}.
As we have reported previously, the sedimented silica spheres repel each other as predicted
by Eq.~(\ref{eq:dlvo}) at the largest inter-plated separations considered, $H = 200~\micron$ \cite{behrens01a,han03a}.
Reducing $H$ does not perceptibly change the spheres' equilibrium height $h$ above the 
lower glass wall, yet nonetheless introduces a minimum into $u(r)$ consistent with a 
long-ranged attraction \cite{han03a}.
The example in Fig.~\ref{fig:ur} at $H = 9~\micron$ has a minimum roughly $0.3~k_B T$ deep
at $r = 1.5 \, \sigma$.
Well-resolved minima are evident for spacings as large as $H \leq 30~\micron$.

Even the small amount of scatter in the measured pair potentials
yields unacceptably large fluctuations in the derivatives used to calculate
the configurational temperatures.
We avoid this by fitting the experimental data to fifth-order polynomials, 
as shown in Fig.~\ref{fig:ur}, and using the fits
as inputs to Eqs.~(\ref{eq:thyper}), (\ref{eq:tcon1}) and (\ref{eq:tcon2}).

\begin{figure}[t!]
  \centering
  \includegraphics[width=\columnwidth]{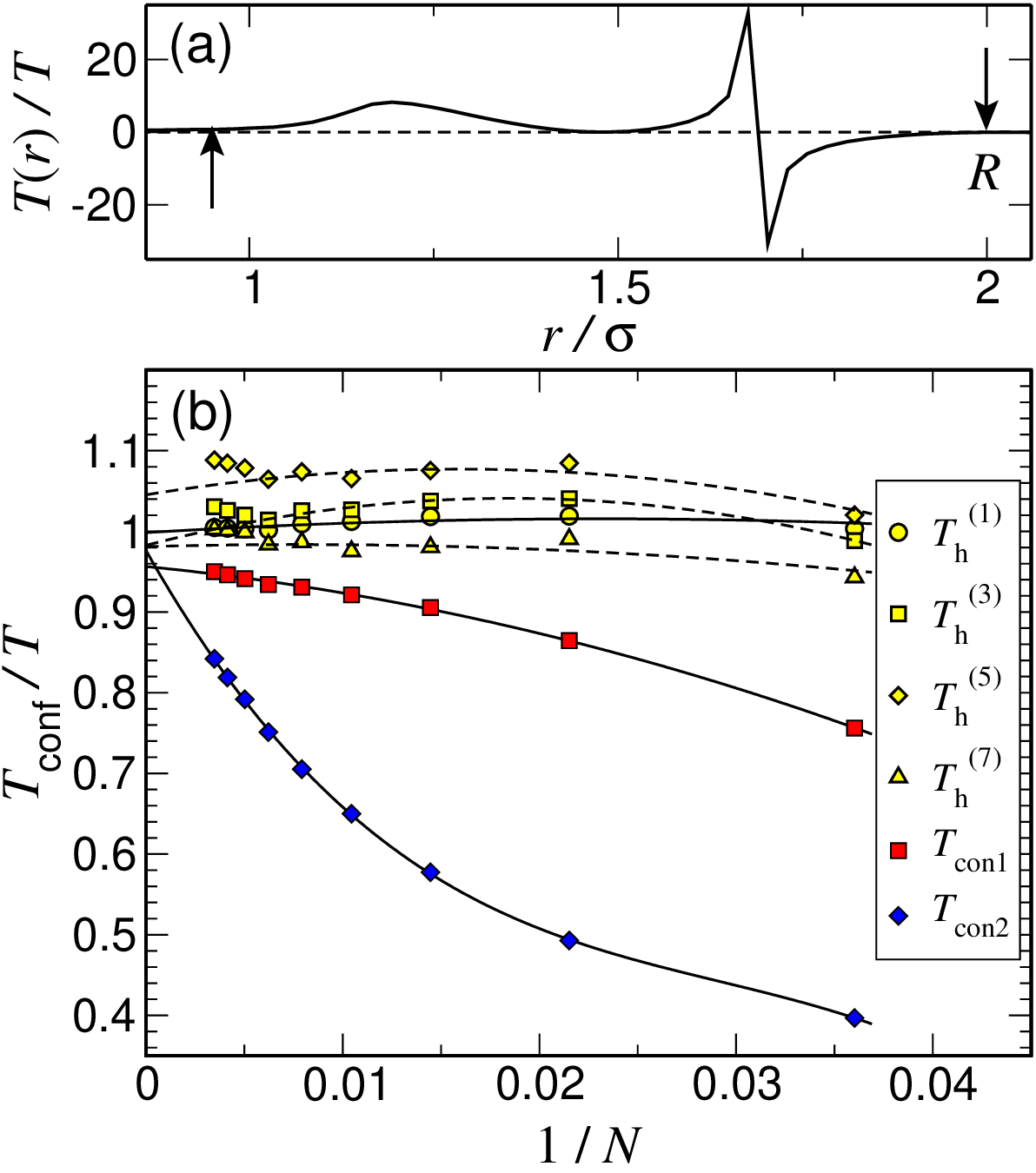}
  \caption{(a) Estimating the interaction range for the $H = 9~\micron$ data
    set using $T(r)$.  Configurational temperatures were calculated over
    $0.93 < r/\sigma < 2$.
    (b) Configurational and hyperconfigurational temperatures as a function of sample size, 
    showing finite-size scaling through fits to second-order polynomials in $1/N$
    (third-order for $\Tcon{2}$).
  }
  \label{fig:tscaling}
\end{figure}

In calculating the total force $\vec{F}_i$ on the $i$-th particle, we must consider
all relevant pair interactions.
Particles close to the edge of the field of view may have strongly interacting
neighbors just out of view.
To avoid errors
due to the resulting spuriously unbalanced forces, we restrict
ensemble averages to particles farther from the edges than the
range $R$ of the interaction.
We estimate $R$ by plotting $T(r)/T = 2\pi (r/\sigma) \, g(r) \, |\nabla u(r)|^2 / \nabla^2 u(r)$,
as shown in Fig.~\ref{fig:tscaling}(a),
which qualitatively gauges contributions to the
configurational temperature due to particles at separation $r$.
This is most useful for systems with weak three-body correlations.
Typically, $R \lesssim 2.0\, \sigma$ for our samples.

Sample imperfections such as a small number of dimers 
also distort the apparent force distribution
We omit such particles from the ensemble averages.
By contrast, we include the small number of 
pair interactions with $r/\sigma \lesssim 1$, which arise from the samples'
polydispersity and also because of projection errors due to out-of-plane
motions for particles near contact.
Omitting these has little effect on \TconF, 
but leads to large systematic errors in higher-order \Ths.

The restricted sample includes only $N = \order{100}$ particles, hardly
the thermodynamic limit.
To correct for finite-size effects, we deliberately
sub-sample our data, and plot the configurational temperature as a function
of $1/N$, as shown in Fig.~\ref{fig:tscaling}(b).
Polynomial fits account for the finite-size
scaling of Eq.~(\ref{eq:general}), and permit
extrapolations to large $N$.
As expected, \Tcon{1} and \Tcon{2} are more sensitive to sample size than \Ths because their derivations
involve ignoring more terms of \order{1/N}.
Despite these differences, all orders of \Ths, as
well as \Tcon{1} and \Tcon{2}, extrapolate to the thermodynamic temperature
in the large-$N$ limit.
This successful outcome strongly suggests that the experimentally
determined potential, including its long-ranged attraction, accurately describes
the dispersion's equilibrium pair interactions.

\begin{figure}[t!]
  \centering
  \includegraphics[width=\columnwidth]{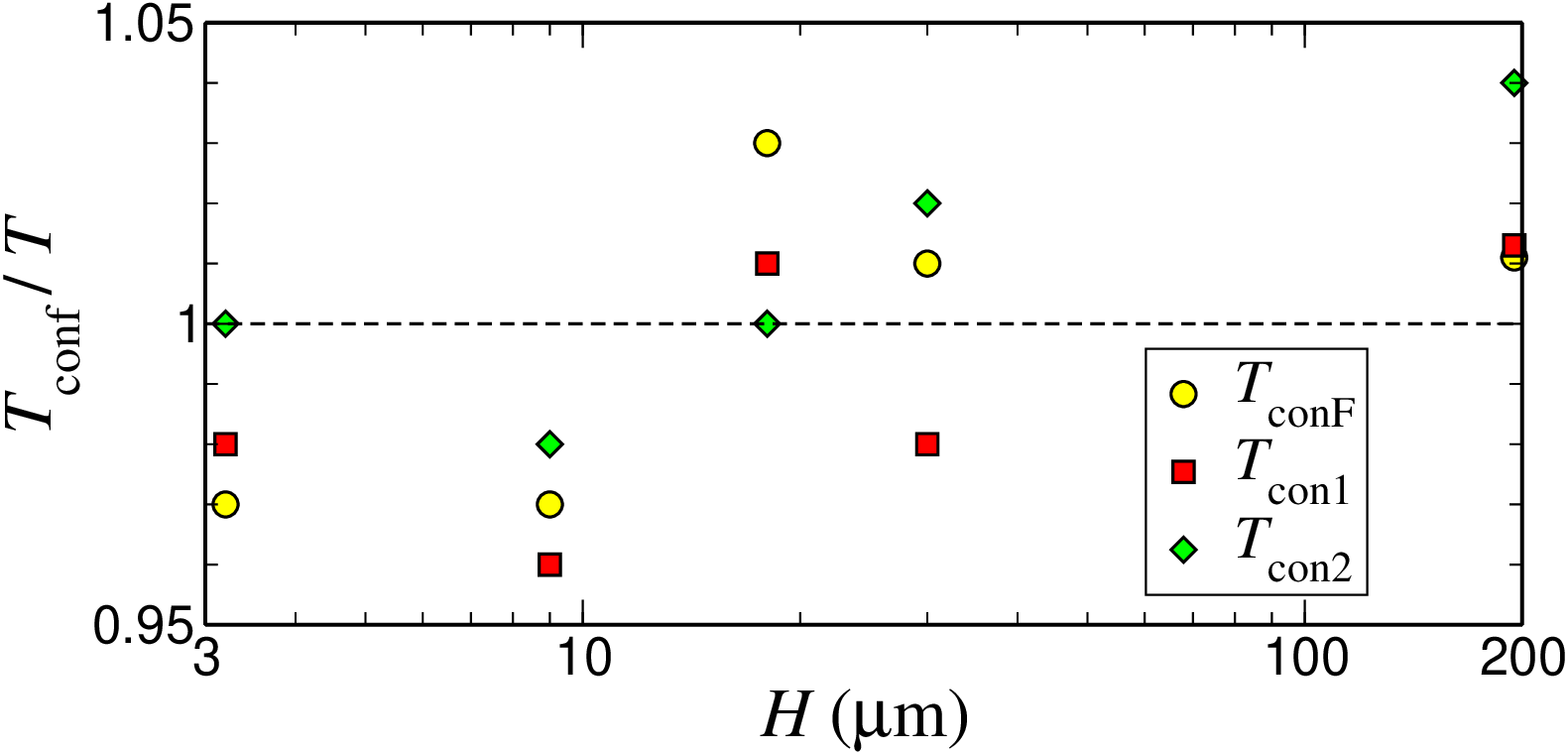}
  \caption{Configurational temperatures at different plate separations. The uncertainty of each 
    extrapolated temperature is less than 0.02.}
  \label{fig:TvsH}
\end{figure}

The configurational temperatures are exceedingly sensitive to small variations
in $\beta u(r)$, with systematic adjustments as small as 0.01 in the short-range repulsive
core leading to variations in the configurational temperature as large as ten percent.
Truncating the attractive part of $u(r)$ increases the configurational temperatures
by fifty percent.
Similarly, failure to precisely correct the imaging system's aspect ratio 
can lead to large deviations.  This potential source of error can be monitored by factoring
\TconF into components along and transverse to the video scan lines and comparing the results.
Deliberately rescaling one axis by 0.01 causes the apparent temperatures along the two directions
to differ systematically by as much as ten percent for all samples.  

Figure~\ref{fig:TvsH} demonstrates that the configurational temperatures
are consistent with the thermodynamic temperature over the entire range of wall separations from
$H = 3.2~\micron$ to $H = 195~\micron$,
despite variations in the form of the associated pair potential \cite{han03a}.
This allows us to draw
several conclusions regarding the nature of confinement-mediated colloidal interactions.
Primarily, we conclude that the measured pair potentials accurately
and self-consistently describe the colloidal particles' interactions. 
To the extent that the configurational temperatures are sensitive to local departures from
equilibrium \cite{ennis01}, the result $T_\text{conf}/T = 1.00 \pm 0.04$ for all of our samples
suggests that anomalous confinement-induced like-charge colloidal attractions cannot be ascribed to
nonequilibrium mechanisms such as hydrodynamic coupling due to transient flows in the
sample \cite{popov02}.

In summary, we have calculated the configurational temperature for experimentally determined
distributions of colloidal silica spheres using their measured pair potentials as inputs.
The success of this procedure provides a thermodynamic self-consistency test for the
measured potentials, and thus establishes that confinement
induces equilibrium pair attractions between charged colloidal spheres.
It furthermore demonstrates the configurational temperature to be a powerful new
tool for experimental condensed matter physics.

We are grateful to Owen Jepps for introducing us to the notion of configurational temperatures,
to Sven Behrens for extensive discussions, and to the donors of the Petroleum Research
Fund of the American Chemical Society for support.

\end{document}